\documentclass[aps,pra, twocolumn, amsmath, amssymb, showpacs, superscriptaddress]{revtex4-2}

\usepackage{graphicx}
\usepackage{bm}

\usepackage{xcolor}

\begin{document}

\title{Theory of composite Ramsey sequences of radiofrequency pulses beyond the rotating wave approximation}

\author{V.\,I.\,Yudin}
\email{viyudin@mail.ru}
\affiliation{Novosibirsk State University, Novosibirsk, 630090 Russia}
\affiliation{Institute of Laser Physics, Siberian Branch, Russian Academy of Sciences, Novosibirsk, 630090 Russia}
\affiliation{Novosibirsk State Technical University, Novosibirsk, 630073 Russia}
\author{O.\,N.\,Prudnikov}
\affiliation{Novosibirsk State University, Novosibirsk, 630090 Russia}
\affiliation{Institute of Laser Physics, Siberian Branch, Russian Academy of Sciences, Novosibirsk, 630090 Russia}
\author{A.\,V.\,Taichenachev}
\affiliation{Novosibirsk State University, Novosibirsk, 630090 Russia}
\affiliation{Institute of Laser Physics, Siberian Branch, Russian Academy of Sciences, Novosibirsk, 630090 Russia}
\author{M.\,Yu.\,Basalaev}
\affiliation{Novosibirsk State University, Novosibirsk, 630090 Russia}
\affiliation{Institute of Laser Physics, Siberian Branch, Russian Academy of Sciences, Novosibirsk, 630090 Russia}
\affiliation{Novosibirsk State Technical University, Novosibirsk, 630073 Russia}
\author{V.\,G.\,Pal'chikov}
\affiliation{All-Russian Research Institute of Physical and Radio Engineering Measurements, Mendeleevo, Moscow region, 141570 Russian}
\affiliation{National Research Nuclear University MEPhI (Moscow Engineering Physics Institute), Moscow, 115409 Russia}
\author{S.\,N.\,Bagayev}
\affiliation{Novosibirsk State University, Novosibirsk, 630090 Russia}
\affiliation{Institute of Laser Physics, Siberian Branch, Russian Academy of Sciences, Novosibirsk, 630090 Russia}

\begin{abstract}
We develop a theory of composite Ramsey sequences of rf pulses interacting with the Zeeman structure at the long-lived atomic level, beyond the rotating wave approximation.  Such sequences are proposed in experiments to detect the violation of local Lorentz invariance [R. Shaniv, et al., Phys. Rev. Lett. \textbf{120}, 103202 (2018)]. Based on Fourier analysis, we have shown that taking into account non-resonant contributions leads to a radical change in the dynamics of the quantum system (with respect to the rotating wave approximation) in the case when the number of Ramsey pulses exceeds several tens. As a result, the effectiveness of using such rf pulses sequences to test local Lorentz invariance has not yet been fully determined and requires additional research.
\end{abstract}

\maketitle

The independence of the results of any local experiment from the velocity and spatial orientation of the measuring instruments is theoretically formulated as local Lorentz invariance (LLI), which is one of the main fundamental principles of symmetry in modern physics. However, some theories that unify the Standard Model and gravity in a single quantum-consistent theory suggest possible breaking of Lorentz symmetry at the Planck scale \cite{Horava_2009,Pospelov_2012,Cognola_2016}.

From the viewpoint of detecting a hypothetical LLI violation, precision spectroscopic experiments with trapped atoms and ions are currently one of the most promising directions of research \cite{Pruttivarasin_2015, Dzuba_2016, Safronova_2018, Megidish_2019, Sanner_2019, Shaniv_2018, Dreissen_2022}. For example, in Ref.\,\cite{Shaniv_2018}, atoms (ions), located at a long-lived energy level with angular momentum $J>1/2$, are considered in the presence of a static magnetic field that removes degeneracy in Zeeman sublevels with a quantum number $m$ ($ -J\leq m\leq J$). In this case, the LLI violation leads to an additional energy shift of Zeeman sublevels according to the law $\kappa m^2$, which can be detected during long-term interrogation of atoms by Ramsey sequences of rf pulses. In this case, the daily variation of the measured coefficient $\kappa$ (due to the rotation of the Earth, with other laboratory conditions remaining unchanged) can be a criterion for LLI violation.

To suppress the influence of fluctuations of static and rf magnetic fields, while maintaining high sensitivity to the small tensor shift ($\propto m^2$), in Ref.\,\cite{Shaniv_2018} it was proposed to use the dynamical decoupling method \cite {Viola_1999,Lidar_2005,Lidar_2007,Khodjasteh_2013,Lidar_2013,Genov_2017}, when the  Ramsey sequence consists of a large number of individual pulses (in our case, hundreds and thousands), the phases of which are also individual and follow a certain law. In relation to LLI testing, a similar technique was applied in the experiment \cite{Dreissen_2022}, where the Ramsey sequence has of several thousand pulses. It should be noted that the effectiveness of the dynamic decoupling technique was rigorously justified within the framework of the two-level atom model and the rotating wave approximation, which is absolutely adequate in the case of optical transitions with frequencies higher than $10^{13}$\;Hz. However, since the frequency of Zeeman splitting in Refs.\,\cite{Shaniv_2018,Dreissen_2022} does not exceed 10\;MHz, then in this case we cannot automatically assume that the effectiveness of this technique will remain unchanged for composite Ramsey sequences of rf pulses. Therefore, for a more detailed development of the method \cite{Shaniv_2018}, the theoretical analysis should be carried out without using the rotating wave approximation.

In this letter, based on Fourier analysis beyond the rotating wave approximation, we have numerically studied the spectroscopic scheme proposed in Ref.\,\cite{Shaniv_2018}. It is shown that taking into account non-resonant contributions leads to a radical change in the dynamics of the quantum system (with respect to the rotating wave approximation used in \cite{Shaniv_2018}) in the case when the number of rf Ramsey pulses exceeds several tens.

Let us consider an atom (ion) at a long-lived energy level with angular momentum $J$, the wave function of which we will describe in the basis of magnetic (Zeeman) sublevels $|J,m\rangle$ (where $-J\,$$\leq$$\,m\,$$\leq$$\,J$). The method of Ramsey sequences of rf pulses is based on the presence of static magnetic field ${\bf B}_{\rm st}=B_{\rm st}{\bf n}_{\rm st}$ and time-modulated field ${\bf B}_{\rm rf}(t)=B_{\rm rf}(t){\bf n}_{\rm rf}$, where the unit vectors ${\bf n}_{\rm st}$ and ${\bf n}_{\rm rf}$ describe the orientation of the static and rf fields, respectively. For a series of pulses with harmonic modulation at the frequency $\nu$, the scalar amplitude of rf magnetic field is described as follows:
\begin{equation}\label{B_rf}
B_{\rm rf}(t)=\eta(t){\cal B}\cos[\nu t +\phi(t)],
\end{equation}
where $\eta(t)=1$ during the Ramsey pulse and $\eta(t)=0$ during the free intervals between pulses, ${\cal B}$ is the field amplitude during the rf pulses, $\phi(t)$ is the phase of the pulse, which can be different for different pulses. In this case, the Hamiltonian of the atom has the following form:
\begin{equation}\label{H}
\widehat{H}(t)=\Omega^{}_{L}(\hat{{\bf J}}\cdot{\bf n}_{\rm st})+\eta(t)2\Omega_{\rm rf}\cos[\nu t +\phi(t)](\hat{{\bf J}}\cdot{\bf n}_{\rm rf})\,,
\end{equation}
where $\hat{{\bf J}}$ is the angular momentum operator, $\Omega^{}_{L}=\mu^{}_{B}g^{}_{J}B_{\rm st }/\hbar$ is the Larmor frequency in a static magnetic field ($\mu^{}_{B}$ is the Bohr magneton, $g^{}_{J}$ is the $g$-factor of the energy level under consideration), and $\Omega_{\rm rf}=\mu^{}_{B}g^{}_{J}{\cal B}/(2\hbar)$ is the effective Rabi frequency for rf field.

In the case of mutually orthogonal orientation of vectors (${\bf n}_{\rm st}$$\,\perp\,$${\bf n}_{\rm rf}$), we choose the $Oz$ axis along ${\bf n}_{\rm st}$ and the $Ox$ axis along ${\bf n}_{\rm rf}$, which leads to the following expression
\begin{equation}\label{H1}
\widehat{H}(t)=\Omega^{}_{L} \hat{J}^{}_{z}+\kappa\hat{J}^{2}_{z}+\eta(t)2\Omega_{\rm rf}\cos[\nu t +\phi(t)]\hat{J}^{}_{x}\,,
\end{equation}
where we also introduce a small tensorial shift $\kappa\hat{J}^{2}_{z}$ ($\propto m^2$) due to the hypothetical violation of LLI, as well as a second-order Zeeman shift and the electric quadrupole shift (originating in ion traps from their inherent electric field gradient) (see Ref.\,\cite{Shaniv_2018}). In this case, the wave function $|\Psi(t)\rangle$, satisfying the Schr\"{o}dinger equation
\begin{equation}\label{Sch}
i\hbar\partial^{}_{t}|\Psi(t)\rangle=\widehat{H}(t)|\Psi(t)\rangle,\;\; |\Psi(t)\rangle=\sum_{m}a^{}_{m}(t)|J,m\rangle,
\end{equation}
is defined in a standard way as a superposition over Zeeman sublevels.

Using the following transformation in Eq.\,(\ref{Sch})
\begin{equation}\label{a}
a^{}_{m}(t)=\tilde{a}^{}_{m}(t)\,e^{-im\nu t},
\end{equation}
it can be shown that the new wave function $|\widetilde{\Psi}(t)\rangle$ is determined from another Schr\"{o}dinger equation
\begin{equation}\label{Sch_eff}
i\hbar\partial^{}_{t} |\widetilde{\Psi}(t)\rangle=\widehat{H}_{2}(t)|\widetilde{\Psi}(t)\rangle,\; |\widetilde{\Psi}(t)\rangle=\sum_{m}\tilde{a}^{}_{m}(t)|J,m\rangle,
\end{equation}
with modified Hamiltonian
 \begin{align}\label{H_new}
&\widehat{H}_{2}(t)=-\delta \hat{J}^{}_{z}+\kappa\hat{J}^{2}_{z}+\eta(t)\frac{\Omega_{\rm rf}}{2}\big\{e^{i\phi(t)}\hat{J}^{}_{-}+e^{-i\phi(t)}\hat{J}^{}_{+}\big\}+\nonumber\\
&\qquad \quad\eta(t)\frac{\Omega_{\rm rf}}{2}\big\{e^{-i2\nu t-i\phi(t)}\hat{J}^{}_{-}+e^{i2\nu t+i\phi(t)}\hat{J}^{}_{+}\big\},
\end{align}
where $\delta$$\,=\,$$\nu-\Omega^{}_{L}$ is the detuning from resonance, and $\hat{J}^{}_{\pm}=\hat{ J}^{}_{x}\pm i\hat{J}^{}_{y}$ are standard up ($+$) and down ($-$) operators.

Further, if the conditions are met
\begin{equation}\label{res_cond}
\frac{|\delta|}{\Omega_{L}}\ll 1,\quad\frac{\Omega_{\rm rf}}{\Omega_{L}}\ll 1,
\end{equation}
one can proceed to the rotating wave approximation by removing the oscillating contributions in Eq.\,(\ref{H_new}) and using only the effective (reduced) resonant Hamiltonian
\begin{align}\label{H_res}
&\widehat{H}_{\rm res}=-\delta \hat{J}^{}_{z}+\kappa\hat{J}^{2}_{z}+\eta(t)\frac{\Omega_{\rm rf}}{2}\big\{e^{i\phi(t)}\hat{J}^{}_{-}+e^{-i\phi(t)}\hat{J}^{}_{+}\big\}=\nonumber\\
&\quad -\delta \hat{J}^{}_{z}+\kappa\hat{J}^{2}_{z}+\eta(t)\Omega_{\rm rf}\big\{\cos\phi(t)\hat{J}^{}_{x}+\sin\phi(t)\hat{J}^{}_{y}\big\}.
\end{align}
It is this resonant approach that was used in Refs.\,\cite{Shaniv_2018,Yeh_2023} to theoretically substantiate the method for LLI testing.

\begin{figure}[t]
\centerline{\scalebox{0.47}{\includegraphics{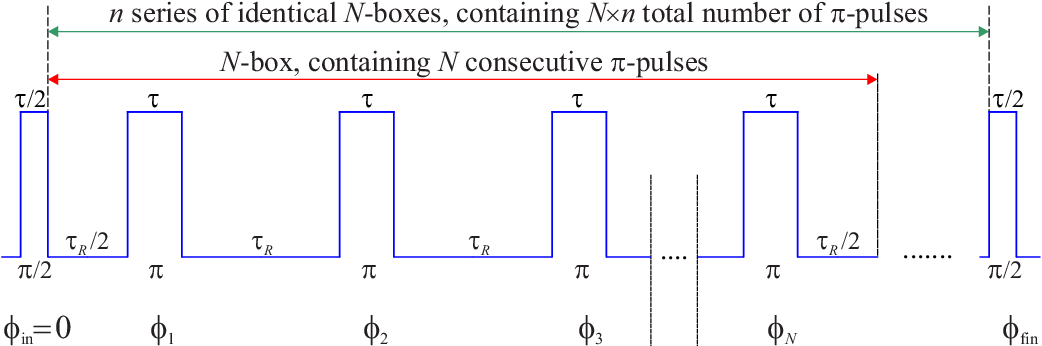}}}\caption{General scheme of a composite Ramsey sequence  \{($\phi^{}_{1},\phi^{}_{2},...,\phi^{}_{N}$)$^n$,\,$\phi^{}_{\rm fin}$\}, consisting of ($Nn+2$) individual Ramsey pulses.} \label{pulses}
\end{figure}

Our purpose is to analyze the applicability of the rotating wave approximation in the case of Ramsey sequences with hundreds and thousands of individual rf pulses. To do this, we carry out numerical calculations within the framework of solving the Schr\"{o}dinger equation with the full Hamiltonian (\ref{H1}) [or (\ref{H_new})] and compare them with calculations for the effective Hamiltonian (\ref{H_res}).

In this paper, we will consider composite Ramsey sequences of rf pulses using the dynamic decoupling technique. The general scheme of such sequences is presented in Fig.\,\ref{pulses}. The initial $\pi/2$-pulse with duration $\tau/2$ (i.e. $\Omega_{\rm rf}\tau/2=\pi/2$) has phase $\phi^{}_ {\rm in}=0$, while the final $\pi/2$-pulse has a phase $\phi^{}_{\rm fin}$. Between these two $\pi/2$-pulses on the time scale, there are $n$ of identical $N$-boxes, each of which consists of $N$ number of $\pi$-pulses with duration $\tau$ ($\Omega_ {\rm rf}\tau=\pi$) and with corresponding phases ($\phi^{}_{1},\phi^{}_{2},...,\phi^{}_{N} $). Thus, there is a total number ($Nn+2$) of individual Ramsey pulses (including the initial and final $\pi/2$ pulses). The free evolution time between neighboring $\pi$-pulses is equal to $\tau^{}_{R}$, while the time between the initial $\pi/2$-pulse and the first $\pi$-pulse, as well as the time between the last $\pi$-pulse and the final $\pi/2$-pulse is equal to $\tau^{}_{R}/2$. The composite Ramsey sequence of rf pulses described above (see Fig.\,1) we will denote as \{($\phi^{}_{1},\phi^{}_{2},...,\phi^{}_{N}$)$^n$,\,$\phi^{}_{\rm fin}$\}.

\begin{figure}[t]
\centerline{\scalebox{0.8}{\includegraphics{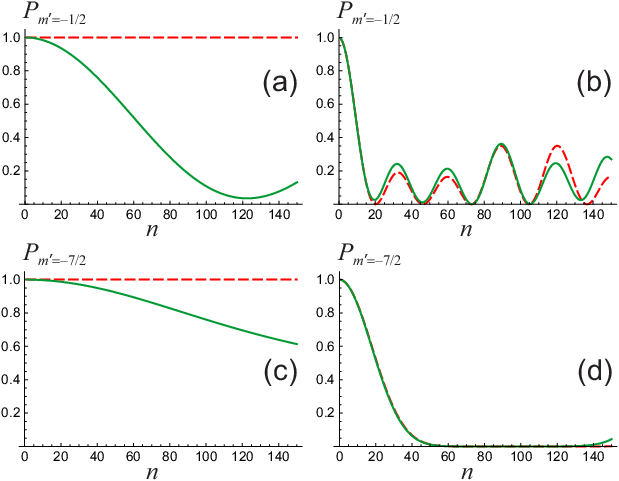}}}\caption{Calculations of the value $P_{m'}$ after the action of the composite Ramsey sequence \{($\pi/2,-\pi/2$)$^n$,\,$\pi$\} for $\kappa=0$ (red dashed line) and $\hbar\kappa/\Omega_{\rm rf}=0.00004$ (green solid line): (a)\;$P_{m'=-\frac{1}{2}}$ based on the resonant Hamiltonian (\ref{H_res}); (b)\;$P_{m'=-\frac{1}{2}}$ based on the full Hamiltonian (\ref{H1}) [or (\ref{H_new})] using Fourier analysis; (c)\;$P_{m'=-\frac{7}{2}}$ based on the resonant Hamiltonian (\ref{H_res}); (d)\;$P_{m'=-\frac{7}{2}}$ based on the full Hamiltonian (\ref{H1}) [or (\ref{H_new})] using Fourier analysis.\\
Calculation parameters: $\Omega_{\rm rf}\tau=\pi$, $\Omega_{L}/\Omega_{\rm rf}=100$, $\delta=0$, $\tau^{}_{R}=10\tau$.} \label{seq1}
\end{figure}

For definiteness, let us consider the energy level with angular momentum $J=7/2$. Such a long-lived (metastable) level ($^{2}$$F$$^{}_{7/2}$) exists in the energy structure of $^{171}$Yb$^{+}$, which has being considered in Refs.\,\cite{Shaniv_2018,Dreissen_2022,Yeh_2023} as one of the most promising objects to test LLI violations. In the calculations, we start with initializing our spin state in Zeeman sublevel with a quantum number $m'$, and after composite Ramsey sequence  \{($\phi^{}_{1},\phi^{}_{2},...,\phi^{}_{N}$)$^n$,\,$\phi^{}_{\rm fin}$\} we analyze the population at the same sublevel, $P_{m'}=|a_{m'}|^2=|\tilde{a}_{m'}|^2$.

\begin{figure}[t]
\centerline{\scalebox{0.8}{\includegraphics{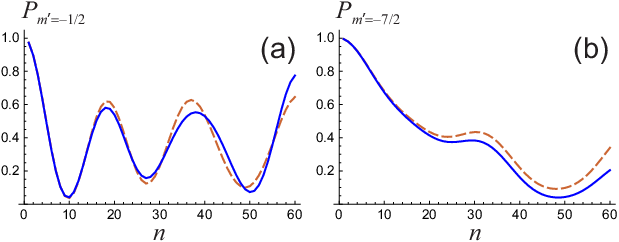}}}\caption{Calculations of the value $P_{m'}$ after the action of the composite Ramsey sequence \{($\pi/2,-\pi/2$)$^n$,\,$\pi$\} in the case when the condition (\ref{neq}) is violated, based on the resonant Hamiltonian (\ref{H_res}) (dashed brown line) and an exact calculation based on the full Hamiltonian (\ref{H1}) [or (\ref{H_new})] (solid blue line): (a)\;for $m'=-1/2$; (b)\;for $m'=-7/2$.\\
Calculation parameters: $\Omega_{\rm rf}\tau=\pi$, $\hbar\kappa/\Omega_{\rm rf}=0.0005$, $\Omega_{L}/\Omega_{\rm rf}=400$, $\delta=0$, $\tau^{}_{R}=10\tau$.} \label{seq3}
\end{figure}

\begin{figure}[t]
\centerline{\scalebox{0.8}{\includegraphics{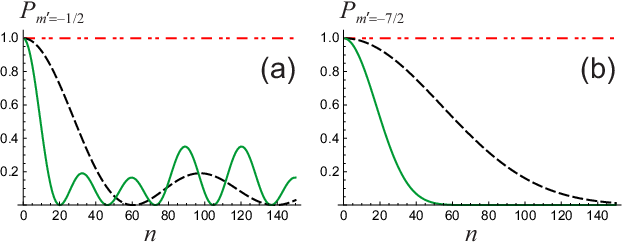}}}\caption{Calculations of the value $P_{m'}$ after the action of the composite Ramsey sequence \{($\pi/2,-\pi/2$)$^n$,\,$\pi$\} based on the resonant Hamiltonian (\ref {H_res}) (dashed-dotted red line), based on the modified resonant Hamiltonian (\ref{H_mod}) (dashed black line), and an exact calculation based on the full Hamiltonian (\ref{H1}) [or (\ref{H_new})] (solid green line): (a)\;for $m'=-1/2$; (b)\;for $m'=-7/2$.\\
Calculation parameters: $\Omega_{\rm rf}\tau=\pi$, $\kappa=0$, $\Omega_{L}/\Omega_{\rm rf}=100$, $\delta=0$, $\tau^{}_{R}=10\tau$.} \label{compar}
\end{figure}

As an example, we carried out calculations for the Ramsey sequence \{($\pi/2,-\pi/2$)$^n$,\,$\pi$\} from Ref.\,\cite{Shaniv_2018}, for which the $N$-box consists of two $\pi$-pulses with phases $\phi^{}_{1}=\pi/2$ and $\phi^{}_{2}=-\pi /2$, and the final $\pi/2$-pulse has the phase $\phi^{}_{\rm fin}=\pi$. Figs.\,\ref{seq1}(a)-(b) show the dependence of $P_{m'=-\frac{1}{2}}$ on the number of $n$ identical $N$-boxes for magnetic quantum number $m'=-1/2$. In this case, Fig.\,\ref{seq1}(a) corresponds to the rotating wave approximation with the effective Hamiltonian (\ref{H_res}) and coincides with the calculations in Ref.\,\cite{Shaniv_2018}, while the calculations in Fig.\,\ref{seq1}(b) are made based on the full Hamiltonian (\ref{H1}) [or (\ref{H_new})], using Fourier analysis. In the latter case, we expand the wave function in a finite number of harmonics $(2{\cal F}+1)$ with numbers from $-{\cal F}$ to ${\cal F}$. The calculations are considered completed when a further increase in ${\cal F}$ does not lead to any noticeable change in the calculated dependencies (for the curves in Figs.\,(\ref{seq1})-(\ref{compar}), the value ${\cal F}=50$ was enough with a good margin).

It should be noted that the main idea of the method \cite{Shaniv_2018}, proposed for the experimental test of LLI violation, is arising from significant influence of the weak tensor shift $\kappa\hat{J}^2_z$ on the long-term dynamics, which is described by the resonant Hamiltonian (\ref{H_res}) [like the green curve in Fig.\,\ref{seq1}(a)]. However, as can be seen from a comparison of Figs.\,\ref{seq1}(a) and (b), the rotating wave approximation very unsatisfactorily describes the atomic dynamics for $n\gg 1$. Moreover, our calculations for the full Hamiltonian (\ref{H1}) [or (\ref{H_new})] show that the small tensor contribution ($\kappa\hat{J}^2_z$) has practically no noticeable effect if the condition is met:
\begin{equation}\label{neq}
\frac{\hbar\kappa}{\Omega_{\rm rf}}\ll\frac{\tau}{\tau^{}_{R}}\frac{\Omega_{\rm rf}}{\Omega_{L}}\,.
\end{equation}
This fact is also clearly confirmed by Figs.\,\ref{seq1}(c)-(d), which show the dependence of $P_{m'=-\frac{7}{2}}$ on the number $n$ of identical $N$-boxes. Similar radical discrepancies between the resonant approach using the effective Hamiltonian (\ref{H_res}) and calculations based on the full Hamiltonian (\ref{H1}) [or (\ref{H_new})] are also observed for other values of $m'$ and $J$ (for example, $J$\,=\,1/2;\;3/2;\;5/2).

If the condition (\ref{neq}) is violated, then exact calculations based on the full Hamiltonian differ little from the resonant approach (see Fig.\,\ref{seq3}). Therefore, a fairly good agreement in \cite{Shaniv_2018} the experimental results for the $^{88}$Sr$^{+}$ ion ($J$\,=\,5/2, $m'$\,=\,$ -3/2$, for sequences from $n$\,=\,1 to $n$\,=\,55) with calculations based on the resonant Hamiltonian (\ref{H_res}) is explained by the fact that the controlled tensorial shift (quadrupole shift of the linear Paul trap) is large enough and the condition (\ref{neq}) is not satisfied. Indeed, our calculations for $J=5/2$, using experimental data from the Supplemental  materials to Ref.\,\cite{Shaniv_2018}, confirm a fairly good agreement between the rotating wave approximation and exact calculations. However, the contribution due to the hypothetical LLI violation will obviously be very small and with the guaranteed fulfillment of the inequality (\ref{neq}) for real experiments with $n\gg 1$. Thus, if the tensorial shift $\kappa\hat{J}^{2}_{z}$ is mainly determined by LLI violation, then the use of rf Ramsey sequence \{($\pi/2,-\pi/2$)$^n$,\,$\phi^{}_{\rm fin}$\}, considered in Ref.\,\cite{Shaniv_2018}, will not allow to reliably detect this shift, even if it exists in reality.

The explanation why for $n\gg 1$ the resonant approach becomes inadequate [if condition (\ref{neq}) is satisfied] is formulated at a qualitative level as follows. For example, let the ratio $(\Omega_{\rm rf}/\Omega_{L})\sim 0.01$ holds. Then, as follows from the Fourier analysis of the Schr\"{o}dinger equation (\ref{Sch_eff}) with the Hamiltonian (\ref{H_new}), the contribution of higher harmonics to the wave function $|\widetilde{\Psi}(t)\rangle$ with respect to the zero harmonic (which is the essence of the rotating wave approximation) is also 0.01 in order of magnitude. Thus, for one Ramsey $\pi$-pulse, an exact calculation gives a deviation of the order of 1\% from the resonant approach. Therefore, after several tens of consecutive $\pi$-pulses, one can expect a significant discrepancy between the rotating wave approximation and the exact calculation, which is what we actually observe (see Fig.\,\ref{seq1}). Similarly, for the ratio $(\Omega_{\rm rf}/\Omega_{L})\sim 0.001$, a significant deviation from the resonant approach can be expected after several hundred Ramsey pulses (which is also confirmed by calculations). At the same time, it is very problematic to significantly reduce the ``parameter of non-resonant contributions'' $(\Omega_{\rm rf}/\Omega_{L})$ so as to achieve the fulfillment of condition (\ref{neq}). Indeed, by greatly increasing the Zeeman splitting $\Omega_{L}$ (i.e. the static magnetic field $B_{\rm st}$) we automatically increase the second-order Zeeman shift, which has the same tensor structure ($\propto \hat{ J}^2_z$) as the supposed shift due to LLI violation. In addition, a noticeable uncontrolled variation in the detuning of $\delta$ becomes possible. A strong decrease in the rf field is also unacceptable, since large relative fluctuations of small magnitude $\Omega_{\rm rf}$ will occur, which will significantly reduce the sensitivity of the method \cite{Shaniv_2018}.

The above circumstances fundamentally distinguish Ramsey sequences of rf pulses from Ramsey sequences for optical transitions, where the ``parameter of non-resonant contributions'' can be extremely small, $(\Omega_{\rm Rabi}/\omega_0)<10^{-10}$ (where $\Omega_{\rm Rabi}$ is the Rabi frequency of the probe field at an optical transition with frequency $\omega_0$). At the same time, the dynamical decoupling method \cite{Viola_1999,Lidar_2005,Lidar_2007,Khodjasteh_2013,Lidar_2013,Genov_2017} has been theoretically justified within only the rotating wave approximation for the two-level atom. Therefore, the effectiveness of this technique for composite Ramsey sequences of rf pulses should be investigated separately outside the framework of the resonant approach.

It is also well known that the first correction to the rotating wave approximation describes a frequency shift caused by non-resonant oscillating terms in the full Hamiltonian (Bloch-Siegert shift \cite{Bloch_1940}). In our case, considering the full Hamiltonian (\ref{H_new}) under the conditions (\ref{res_cond}), it can be easily shown that taking into account the first non-resonant contributions leads to the modified resonant Hamiltonian
\begin{equation}\label{H_mod}
\widehat{H}^{\rm (mod)}_{\rm res}=\widehat{H}_{\rm res}+\eta(t)\frac{\Omega^2_{\rm rf}}{4\Omega_{L}}\hat{J}^{}_{z} ,
\end{equation}
where we used $\eta^2(t)=\eta(t)$. Thus, the expression for $\widehat{H}^{\rm (mod)}_{\rm res}$ differs from (\ref{H_res}) by additional small shifts in the energy of Zeeman sublevels ($\propto\hat{J}^{}_{z}$) during the action of pulses ($\eta(t)=1$), while during the free evolution time these shifts are absent ($\eta(t)=0$). However, as shown in Fig.\,\ref{compar}, even the use of the modified resonant approximation based on the Hamiltonian (\ref{H_mod}) does not lead to an adequate description of the atomic dynamics for composite sequence of rf pulses with $n\gg 1$.

In conclusion, we have considered the composite Ramsey sequences of rf pulses interacting with Zeeman structure of the long-lived atomic level using Fourier analysis beyond the rotating wave approximation. Such sequences were proposed in Ref.\,\cite{Shaniv_2018} for experimental testing of LLI. We have shown that taking into account non-resonant contributions leads to a radical change in the dynamics of the quantum system (with respect to the rotating wave approximation) in the case when the number of Ramsey pulses exceeds several tens. This is explained by the fact that the ``parameter of non-resonant contributions'' (the ratio $\Omega_{\rm rf}/\Omega_{L}$) is not very small ($\Omega_{\rm rf}/\Omega_{L}\,$$\sim$\,10$^{-2}$-10$^{-3}$). As a result, the possibility of using such Ramsey sequences of rf pulses to detect LLI violation is not fully determined and requires additional researches of various spectroscopic schemes. At the same time, the effectiveness of the dynamical decoupling technique (i.e., the effectiveness of suppressing the influence of magnetic fields fluctuations, in our case) is also not guaranteed and should be studied separately beyond the rotating wave approximation.

The obtained results, in addition to the development of methods for LLI testing, can also be important for the theoretical analysis in various branches of quantum metrology that use degenerated energy levels of atoms (ions) and fields with rf modulation (for example, in atomic clocks \cite{Zanon_2012}).

We thank T.\,E.\,Mehlst\"{a}ubler, L.\,S.\,Dreissen, Ch.-H.\,Yeh, H.\,A.\,F\"{u}rst, and K.\,C.\,Grensemann for useful discussions.

\end{document}